\documentclass[preprint,authoryear,12pt]{elsarticle} 
\usepackage{amsthm}
\usepackage[ps2pdf,
a4paper=true,
breaklinks=true,
colorlinks=true,
pdfauthor={First Author et al.},
pdftitle={Template for manuscripts in Advances in Space Research}
]{hyperref}
\biboptions{comma,round}
\journal{Advances in Space Research}
\begin{document}
\begin{frontmatter}
\title{S5 0716+714 : GeV variability study  }
\author[label1,label2]{B. Rani\corref{cor}}
\ead{brani@mpifr-bonn.mpg.de}
\cortext[cor]{Corresponding author}
\author[label1]{T. P.\ Krichbaum}
\author[label3]{B. Lott}
\author[label1]{L. Fuhrmann}
\author[label1]{J.A. Zensus } 
\address[label1]{Max-Planck-Institut f$\ddot{u}$r Radioastronomie (MPIfR), Auf dem H{\"u}gel 69, D-53121 Bonn, Germany}
\address[label2]{Aryabhatta Research Institute of Observational Sciences (ARIES), Manora Peak, Nainital, 263 129, India}
\address[label3]{Universit{\'e} Bordeaux 1, CNRS/IN2p3, Centre d'Etudes Nucle{\'a}ires de Bordeaux Gradignan, 33175 Gradignan, France}

\begin{abstract}
The GeV observations by {\it Fermi-LAT} give us the opportunity to characterize
the high-energy emission (100 MeV - 300 GeV) variability properties of the BL Lac object
S5 0716+714. In this study, we performed flux and spectral analysis of more than 3 year
long (August 2008 to April 2012) {\it Fermi-LAT} data of the source. During this period, the source exhibits two 
different modes of flux variability with characteristic timescales of $\sim$75 and $\sim$140 days, respectively. We also 
notice that the flux variations are characterized by a weak spectral hardening.  The GeV spectrum of the source 
shows a clear deviation from a simple power law, and is better explained by a broken power law.  Similar to 
other bright Fermi blazars, the break energy does not vary with the source flux during the 
different activity states.  
We discuss several possible scenarios to explain the observed spectral break. 

\end{abstract}

\begin{keyword}
galaxies: active -- BL Lacertae objects: individual: S5 0716+714 --
             Gamma rays


\end{keyword}

\end{frontmatter}


\section{Introduction}
\label{intro}
The BL Lac object S5 0716+714 is an extremely active blazar, which shows significant flux variability on timescales 
from hours to days \citep{raiteri2003, rani2010, gupta2012}. The optical continuum of the source is so featureless that 
it is hard to estimate its redshift.
\citet{nilsson2008} claimed a lower limit of $z = 0.31\pm0.08$ based on the photometric detection of
the host galaxy. Very recently, the detection of intervening Ly$\alpha$ systems in the ultra-violet
spectrum of the source constrains the earlier estimates of $z$ to 0.2315 $< z < $0.3407
\citep{danforth2012}.

S5 0716+714 has been classified as an intermediate-peak blazar (IBL) by \citet{giommi1999}, as the 
frequency of the first spectral energy distribution (SED) peak varies between 10$^{14}$ and 10$^{15}$ Hz, and thus does not 
fall into the wavebands specified by the usual definitions of low and high energy peak blazars 
(i.e. LBLs and HBLs). A concave X-ray spectrum in 0.1-10 keV band adds another factor in 
support of the IBL nature of the source \citep{foschini2006, ferrero2006}. The concave X-ray spectrum 
provides a signature of the presence of both the tail from the synchrotron emission and a flatter part 
from  the Inverse Compton (IC) spectrum.

{\it EGRET} on board the {\it Compton Gamma-ray Observatory (CGRO)} detected high-energy 
$\gamma$-ray ($>$100 MeV) emission of 0716+714 several times during 1991 to 2000 \citep{hartman1999, 
lin1995, nandikotkur2007}. Two strong $\gamma$-ray flares were detected in the source during September and October 
2007 \citep{chen2008}. The broadband spectral modeling suggests the presence of two synchrotron self-Compton 
(SSC) components, representative of a slowly and a rapidly variable component, respectively.

Recently, the {\it MAGIC} collaboration reported the first detection of VHE gamma-rays ($>$100 GeV) from the 
source at a 5.8$\sigma$ significance level \citep{anderhub2009}. The discovery of S5 0716+714 as a VHE 
gamma-ray blazar was triggered by its very high optical state, suggesting a possible correlation 
between the VHE gamma-ray and the optical emission. This source also belongs to the  
{\it Fermi/LAT} Bright AGN Sample (LBAS) \citep{abdo2010LBAS}, where the GeV spectrum of the source is 
described by a broken power law. 
The combined GeV-TeV spectra of the source display deviations from the single power-law, which are  
suggested to be due to absorption in the broad-line region (BLR) in 10--100 GeV energy range 
\citep{senturk2011}.

The GeV observations by {\it Fermi-LAT} give us the opportunity to study and characterize
the high-energy emission (100 MeV - 300 GeV) variability properties of the BL Lac object
S5 0716+714. In this study, we performed flux and spectral analysis of more than 3 year
long (August 2008 to April 2012) {\it Fermi-LAT} data of the source. In this paper, we present 
the results of the LAT observations. This paper is structured as follows. Section 2 provides a brief
description of the observations and data reduction. In Section
3 we report our results, and the discussion is given in Section 4. We conclude our results in Section 5.

\section{Observations and Data Reduction}
\label{lat_data}
The gamma-ray data (100 MeV $-$ 300 GeV) employed here  
are collected over JD = 2454686 (August 08, 2008 [12:00 UTC]) to JD = 2456022 (April 04, 2012 [12:00 UTC]) in 
survey mode by the {\it FERMI/LAT}  instrument. The LAT data are analyzed using the standard 
ScienceTools (software version v9.23.1) and the instrument response function 
P7V6\footnote{http://fermi.gsfc.nasa.gov/ssc/data/analysis/scitools/overview.html}. 
Photons in the source event class ({\it evcls =2}) are selected for this analysis because of
their reduced charged-particle background contamination and a good angular 
reconstruction. A zenith angle $<105^{\circ}$ cut in the instrument coordinates is 
used to avoid gamma-rays from the Earth limb. The diffuse emission from our Galaxy 
is modeled using a spatial model (gal$\_$2yearp7v6$\_$v0.fits), which is refined with the {\it Fermi-LAT} 
data taken during the first two years of operation. The extragalactic diffuse and residual 
instrumental backgrounds are modeled as an isotropic component (isotropic$\_$p7v6source.txt), 
which is provided with the data analysis tools. The data analysis is done with an 
unbinned maximum likelihood technique using the likelihood analysis software developed 
by the LAT team.

We analyzed a Region of Interest (RoI) of 10$^\circ$ in radius, centered 
at the position of the $\gamma$-ray source associated with S5 0716+714, using the 
maximum-likelihood algorithm implemented in {\it gtlike}. In the RoI model, we include all 
the 24 sources within 10$^\circ$ with their model parameters fixed to their catalog values except for 
4C +71.07 (2FGLJ0841.6+7052), as none of the other sources are reported as variable in 2FGL 
catalog \citep[see][for details]{ackermann2011}. 4C +71.07 was reported to be variable source in 
the 2FGL catalog, so we keep all the model parameters free for it. It is important to note that the 
contribution of the other 23 sources within the RoI model to the observed variability of the source is
negligible as they are very faint compared to 0716+714.

The source variability is investigated by producing light curves by likelihood analysis with different  
time binnings (1 day, 1 week and 1 month) and over different energy ranges 
(E $>$ 100 MeV, E $>$ 248 MeV, E = 0.1-1 GeV and E $>$ 1 GeV). The light curves are produced by 
modeling the spectra over each bin using a simple power law  which can provide a good fit over 
these small time bins, since the statistical uncertainties on the power law  indices are smaller than those 
obtained from the broken-power law (BPL) fits. 

The spectral analysis is performed by fitting the GeV spectra with multiple models over the whole energy 
range covered by  {\it Fermi/LAT} above 100 MeV. The different spectral forms are :  simple power law  
[SPL, $N(E) = N_{0} (E)^{\Gamma}$, $N_0$ : Integral flux and $\Gamma$ : photon index], 
and broken power law [BPL, $N(E) = N_0 (E/E_{break})^{-\Gamma_{i}}$, with i = 1 if E $<$ $E_{break}$ 
and i = 2 if E $>$ $E_{break}$, $\Gamma_1$, $\Gamma_2$ : the two photon indices and $E_{Break}$ : break 
energy]. We also examine the spectral behavior over the whole energy range with a SPL 
model fitting over equally spaced logarithmic energy bins with $\Gamma$ kept constant and equal to the value fitted 
over the whole range.

We also computed photon fluxes above the “de-correlation energy” E$_0$ \citep{lott2012} which  minimizes the 
spurious correlations between integrated photon flux and photon index ($\Gamma$).  Over the 
course of 3.8 years of observations, we found $E_0$ = 248 MeV. We follow the adaptive binning analysis method 
\citep{lott2012} to generate the constant uncertainty light curve above $E_0$. 
The estimated systematic uncertainty on the flux is 10$\%$ at 100 MeV, 5$\%$ at 500 MeV, 
and 20$\%$ at 10 GeV.

\section{Results}

\subsection{Sky map}
\begin{figure}
   \centering
\includegraphics[scale=0.33,angle=0]{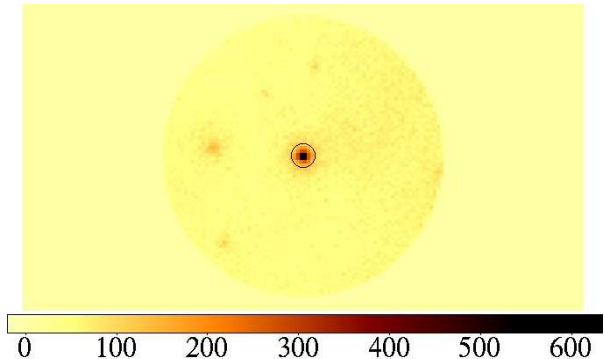}
   \caption{ Sky map of $\gamma$-ray events centered on S5 0716+714 (image radius of 10$^{\circ}$) 
above 100 MeV as measured by {\it Fermi-LAT} over past $>3$ years. The brightness scale at the bottom represents 
the number of observed photons.  
              } 
\label{count_map}
    \end{figure}

Figure \ref{count_map} shows the {\it Fermi-LAT} count map of the $\gamma$-ray events above 
100 MeV centered on the position of S5 0716+714 with an image radius of 10$^{\circ}$. As we see,  
there is no source as bright as 0716+714 within  10$^{\circ}$ of RoI. 
The nominal position of 0716+714 is marked by a circle.
A total of 14,657 $\gamma$-ray photons associated with 0716+714 are detected during $\sim$3.8 years of 
observations within the 68$\%$ containment radius of the LAT PSF above 100 MeV.

\subsection{Temporal behavior}
\label{sec_var}
\subsubsection{Light Curves}
\begin{figure}
   \centering
\includegraphics[scale=0.43,angle=-90]{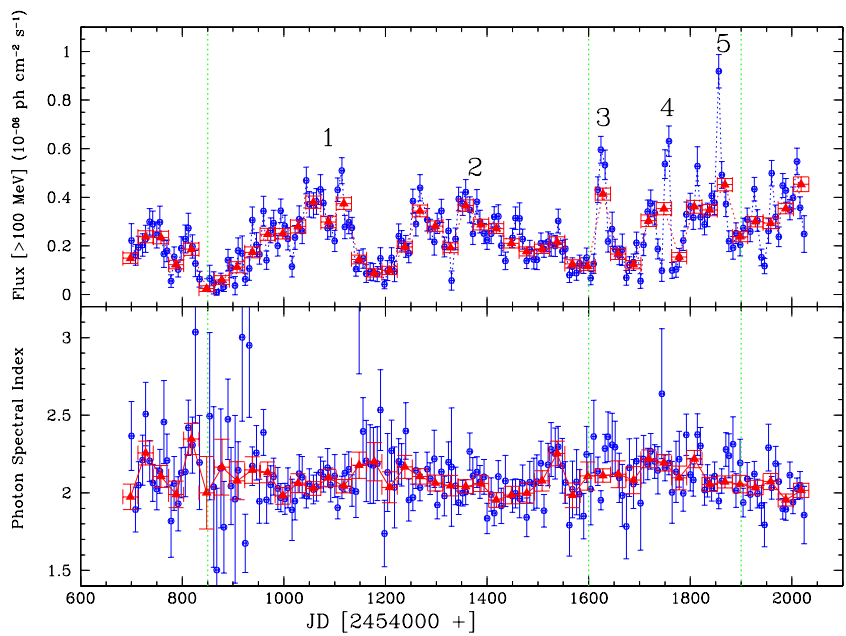}
   \caption{ Gamma-ray flux and photon index light curve of S5 0716+714 measured with the {\it Fermi-LAT} since 
launch till April 04, 2012. The blue symbols show weekly averaged flux while monthly averaged are in 
red. The green lines separate the two different modes of variability observed in the source (see text for details). 
              } 
\label{plot_flx}
    \end{figure}

We investigate the GeV flux variability of 0716+714 over a time period  between August 04, 2008 to 
April 04, 2012. Fig. \ref{plot_flx} shows the weekly and monthly averaged gamma-ray light curves 
extracted over an energy range 100 MeV to 300 GeV. The source displays substantial flux variability 
during the past $\sim$3.8 years of LAT monitoring with five major flares labeled as ``1" to ``5". 
Apparently,  some individual flares are further composed  of a number of sub-flares.

There is a significant enhancement in the weekly 
averaged gamma-ray flux over a time period between JD'\footnote{JD' = JD-2454000 (September 21, 2006)} = 900 to 
1110 (flare 1), peaking at JD' $\sim$ 1110, 
with peak flux equal to (0.57$\pm$0.05)$\times$10$^{-6}$ ph cm$^{-2}$s$^{-1}$, which is  $\sim$6 times brighter than its 
minimum value and $\sim$3 times brighter than its average value. Later it decays reaching a minimum at JD' = 1150 followed  
by a quiescent state until JD' = 1220. The  quiescent state is followed by a low amplitude flux variability (flare 2) and 
later by a sequence of rapid flares (flare 3 to 5).

\begin{figure}
   \centering
\includegraphics[scale=0.32]{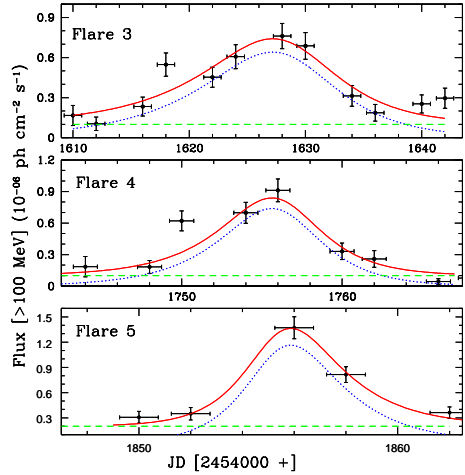}
   \caption{ Light curves of the source above 100 MeV with a time binning of 3 day.  
The lines corresponds to the result of fitted components. The dotted curve is the flaring component. 
The dashed line is the background flux level and the solid curve is the total of the two components.     
}
\label{flare_model}
    \end{figure}

The high photon statistics during the rapid flares allows us to investigate their evolution with a fine time resolution. 
The light curves for $F_{100}$ with a 3 day time binning for the individual flares are shown in Fig. \ref{flare_model}.
A fit consisting of a slowly varying background and sub-flaring components is performed 
for each individual flare.  The slowly varying background is roughly approximated by a photon flux 
value = 0.40 $\times 10^{-7}$ ph cm$^{-2}$ s$^{-1}$. 
 Each component is fitted by a function of the form : 
\begin{equation}
F(t) = 2~F_0 \big [ e^{(t_0 - t)/T_r} + e^{(t - t_0)/T_f} \big ]^{-1}
\end{equation}
where T$_r$ and T$_f$ are the rising and decay times, respectively, and F$_0$ is the flux at t$_0$ representing
approximately the flare amplitude. The solid curves in Fig. \ref{flare_model} represent the fitted flare components 
and the fitted parameters for each sub-component are given in Table \ref{tab1}.

Flare 3 lasts for a duration of $\sim$12 days (JD' = 1610-1638). The source reaches a peak flux 
value F$_{E>100 MeV}$ = (0.63$\pm$0.11)$\times$10$^{-6}$ ph cm$^{-2}$s$^{-1}$ during this flare with a doubling timescale 
of 4.3 days. This flare is followed by another 
rapid flare (flare 4) which has a duration of $\sim$10 days. During this flare, the source reaches a peak flux value = 
(0.73$\pm$0.12)$\times$10$^{-6}$ ph cm$^{-2}$s$^{-1}$ 
above 100 MeV. Flare 5 is the brightest gamma-ray flare observed in the source  with a peak flux value = 
(1.13$\pm$0.03) $\times$ 10$^{-6}$ ph cm$^{-2}$s$^{-1}$ at E $>$ 100 MeV with a doubling timescale of less than a day. 
This is the fastest recorded GeV flare in the source.

\begin{table}
\scriptsize
\caption{Fitted parameters of the rapid flares}
\begin{tabular}{cccccc} \hline
Flare  & T$_r$         &  T$_f$        & t$_0$         & F$_0$               & Doubling  \\
       & (days)        &  (days)       & JD'    & $10^{-6}$ ph cm$^{-2}$ s$^{-1}$ & Time (days)$^{*}$ \\\hline   
3      & 6.22$\pm$1.02 &4.14$\pm$0.86  &1628.2$\pm$0.2 &0.63$\pm$0.11        & 4.31     \\  
4      & 4.21$\pm$0.66 &2.44$\pm$0.34  &1756.0$\pm$0.1 &0.73$\pm$0.12        & 2.91      \\
5      & 1.29$\pm$0.12 &2.05$\pm$0.18  &1855.5$\pm$0.1 &1.13$\pm$0.03        & 0.89      \\\hline
\end{tabular} \\
doubling time = $T_r \times ln2$ \\
\label{tab1}
\end{table}

\begin{figure}
   \centering
\includegraphics[scale=0.35]{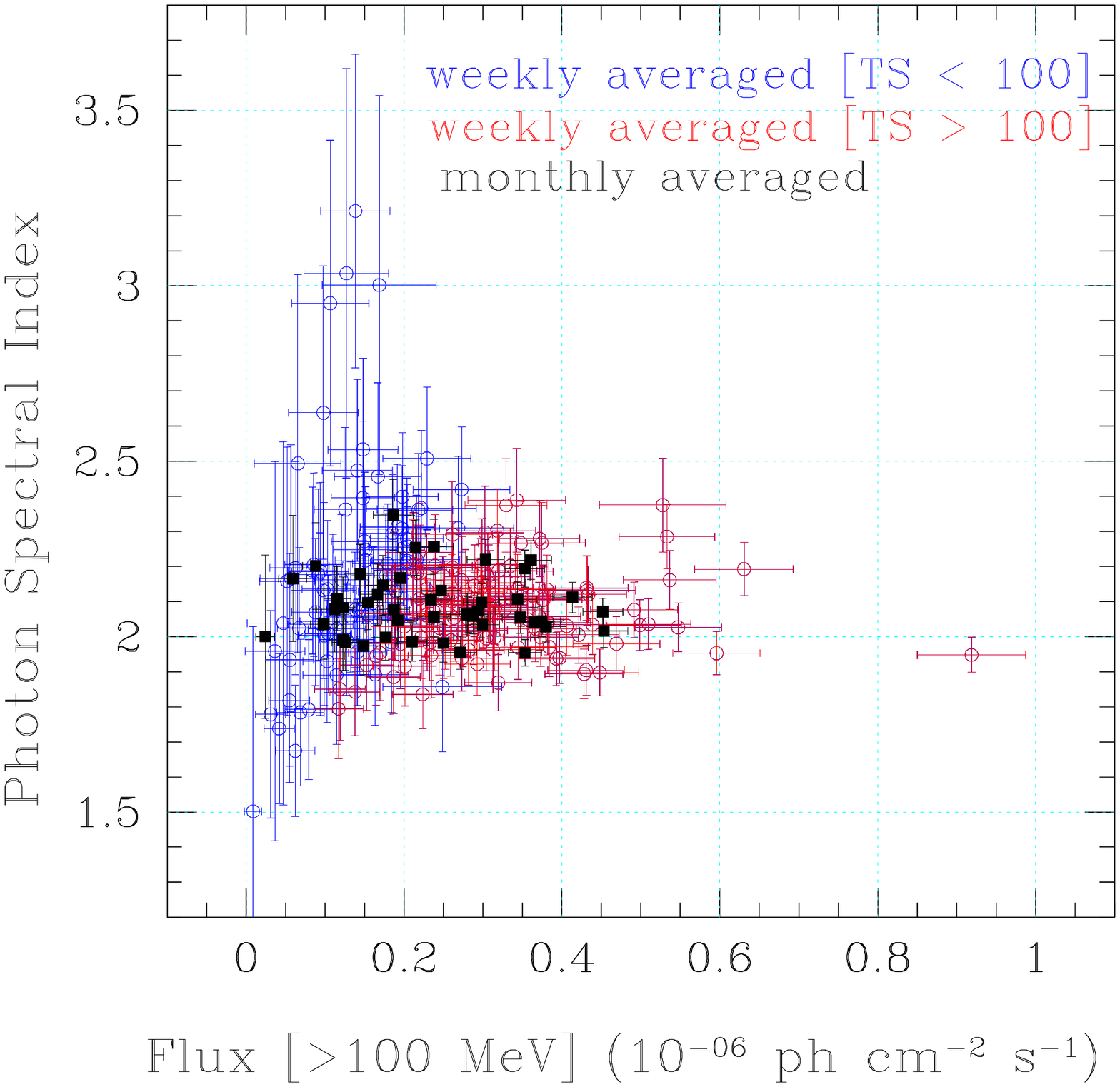}
   \caption{ Photon index ($\Gamma$) vs photon flux variations at E$>$100 MeV. The blue (TS $<$ 100) and 
the red (TS $>$ 100) symbols represent the weekly averaged values while the monthly averaged are in black.       
              }
\label{flx_indx}
    \end{figure}

In comparison to the substantial flux variations, the photon index ($\Gamma$) remains almost constant during  
the different modes of flux activity. We notice only a marginal steepening of spectrum in the monthly averaged light curves 
during the flaring epochs.  
As we see in Fig. \ref{flx_indx}, the photon flux variations are characterized by a weak spectral hardening.  For monthly 
averaged statistics,  $\Gamma$ changes from (2.20$\pm$0.01) to (2.00$\pm$0.04) for a flux variation of (0.10$\pm$0.02) 
to (0.50$\pm$0.01) $\times$ 10$^{-6}$ ph cm$^{-2}$ s$^{-1}$.

\subsubsection{Flux variations at different Energy bands}
\begin{figure*}
\includegraphics[scale=0.8,angle=0]{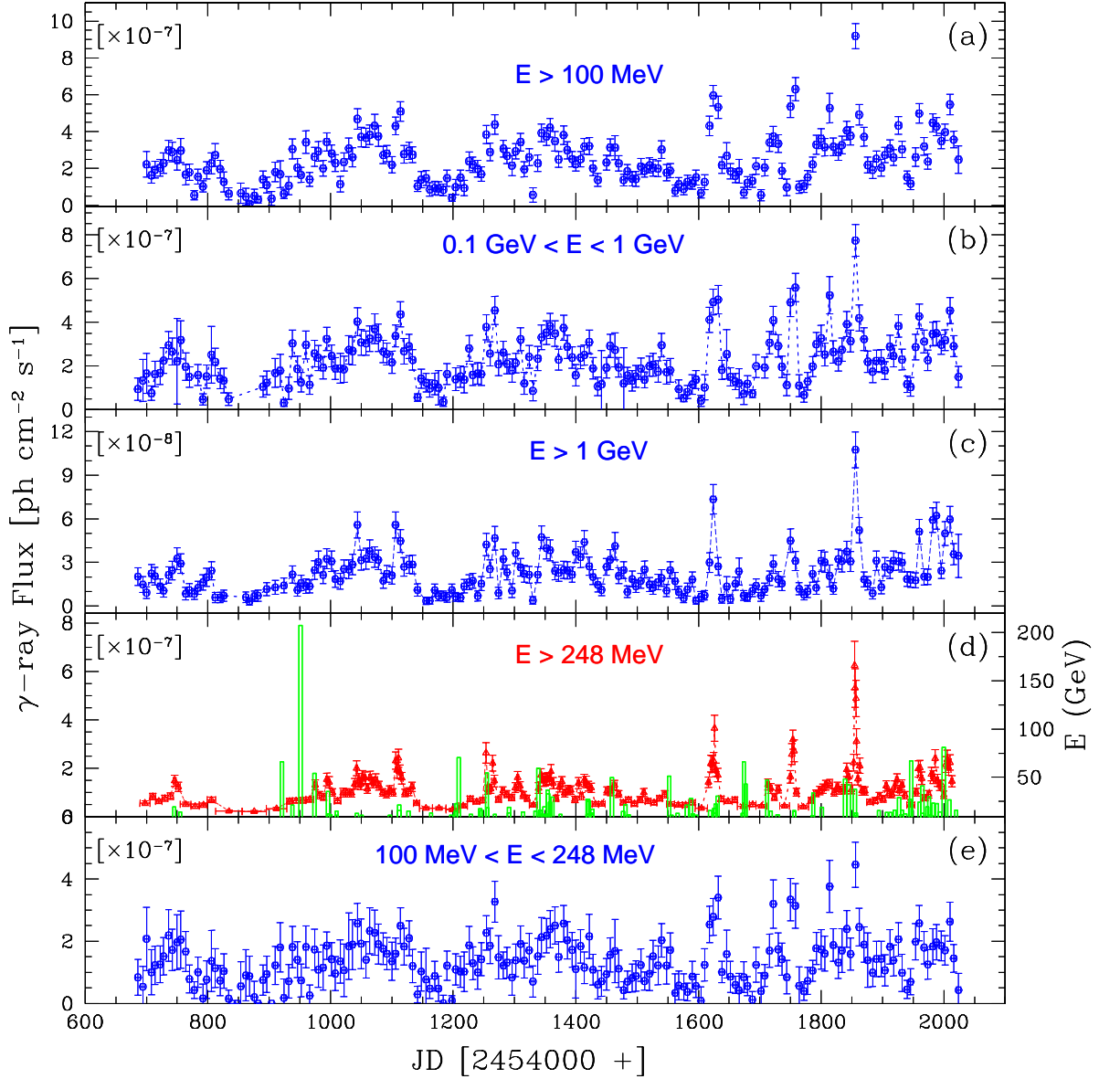}
   \caption{Gamma-ray flux light curves of S5 0716+714 during the first 3.8 years of the {\it Fermi-LAT} observations from 2008 
August to 2012 April; (a) weekly averaged light curve sampled above 100 MeV, (b) weekly averaged light curve sampled 
at 0.1$-$1 GeV, (c) weekly averaged light curve above 1 GeV, (d) the constant uncertainty (15$\%$) light curve above 
de-correlation energy, $E_0 > 248 MeV$ obtained through adaptive binning analysis method. The green histogram represents 
the arrival time distribution of E $>$ 10 GeV photons associated with S5 0716+714 and (e) weekly averaged light curve 
below $E_0 = 248 MeV$.  
       } 
\label{plot_weeklylc}
    \end{figure*}

We have also investigated the temporal characteristics of the source at different energy bands. Fig. \ref{plot_weeklylc} 
shows a comparison of the flux variability at different energies. Figures \ref{plot_weeklylc} (b) \& (c) show the GeV flux 
variations above and below 1 GeV, respectively. We found no substantial difference in the flux variability at E$<$1GeV
and E$>$1GeV. Such a behavior is obvious because of the marginal variation in $\Gamma$. We also do not find any time lag between 
the two light curves (at E below and above 1 GeV) for the weekly averaged light curves. Due to a limited statistics for 
the finer binned light curves, we can not claim any shorter time lag than our binning interval of 7 days.

Fig. \ref{plot_weeklylc} (d) shows the flux variations above the de-correlation energy, $E_0$ = 248 MeV.  
The constant uncertainty (15$\%$) light curve (red symbols) is obtained through the adaptive binning analysis 
method following \citet{lott2012}. An advantage of using 
this method is to avoid upper limits and to obtain better characteristics of the flares. 
The weekly averaged light curve below $E_0$ is shown in 
part (e) of the Fig. \ref{plot_weeklylc}. The variability features are not clearly visible below $E_0$ due to large 
uncertainty and scattering of individual data points.

\subsubsection{Highest energy photons} 
During the 3.8 years of observations, the highest energy photon associated with 0716+714 was detected at
JD = 2454951  with an estimated energy of 207 GeV. 
This photon is observed as a front event of the LAT detector.  The reconstructed arrival direction of the photon
is 0.05$^{\circ}$ away from S5 0716+714, and is within the 68$\%$ containment radius of the LAT PSF at 207 GeV. 
Based on our model fit of the epoch which contains that highest-energy photon, we find the probability that the
photon was associated with S5 0716+714 (as opposed to all other sources in the model including the diffuse emission
and nearby point sources) is 99.96$\%$ which corresponds to 3.56 $\sigma$. 

In total, we found 107 events with estimated energies higher than 10 GeV centered at S5 0716+714 within the 68$\%$ 
confinement radius of the LAT PSF and a total of 10 events above 50 GeV. Fig. \ref{plot_weeklylc} (d) plots the
arrival time distribution of photons above 10 GeV. Interestingly, the highest  energy photon arrived during the 
rising part of flare 1. In fact, a number of several high-energy photons were observed during this period. But during 
the peak and decay of flare 1, the number of events associated with the arrival of high-energy photons is very small. 
For rest of the four flares ``2"-``5", the arrival time distribution of the high-energy photons do not follow any systematic 
trend w.r.t  the photon flux variations (see Fig. \ref{plot_weeklylc} (d)).

\subsubsection{Variability timescale} 
\begin{figure}
   \centering
\includegraphics[scale=0.33,angle=0]{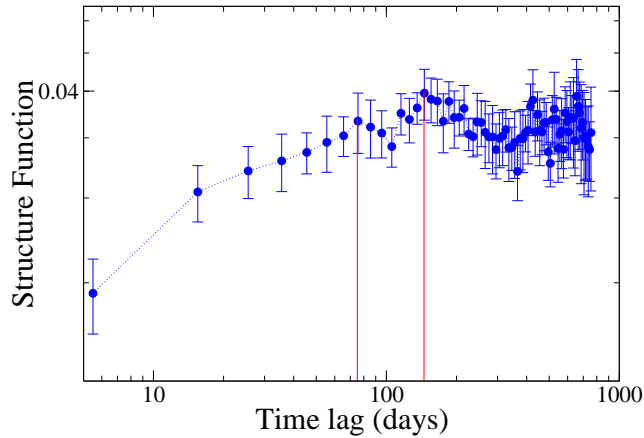}
   \caption{ The $\gamma$-ray structure function analysis curve at E above $>$ 248 MeV with 
a bin size of 10 days.   
               }
\label{plot_sf}
    \end{figure}

In order to extract the characteristic time scale of variability ($t_{var}$) from the GeV 
light curves, we employed the structure function (SF) \citep{simonetti1985} analysis method.  
The formula and details of the method can be found in \citet{rani2009}. 
For the SF analysis, we have used the adaptive binned $\gamma$-ray light curve at E$>$248 MeV. 
The $\gamma$-ray SF curve is shown in Fig. \ref{plot_sf}. 
The SF curve follows a continuous rising trend showing a peak at $\sim$75 days,
followed by another maximum at $\sim$145 days.
So, the SF curve reveals two different variability time scales, one which reflects the short-term variability 
($t_{var1}$) 
while other refers to the long-term variability ($t_{var2}$).
The first SF peak at time lag, $t_{var1}$ = 75$\pm$5 days characterize the 
short-term or fast variability while the second peak at $t_{var2}$ = 140$\pm$5 days represent the 
long-term variability.

\subsection{Spectral behavior}
\label{sec_sed}
\begin{figure}
   \centering
\includegraphics[scale=0.3,angle=-90]{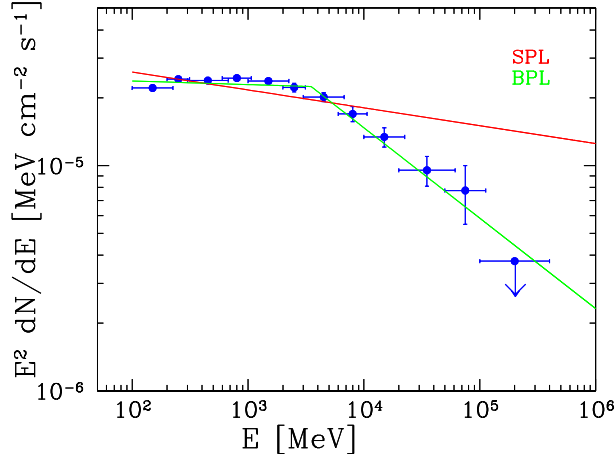}
   \caption{ Spectral energy distribution (SED) of S5 0716+714 during past 
3.8 years of LAT monitoring. The red curve represents the best fitted simple power law, 
while broken power law is in green.    
              } 
\label{plot_spectra_total}
    \end{figure}

We extract the $\gamma$-ray spectrum using data for the entire 3.8 year period. 
Fig. \ref{plot_spectra_total} shows the GeV spectrum of the source with blue symbols as 
spectral measurements over equally spaced logarithmic energy bins in an energy range between 
100 MeV to 300 GeV. The solid curves represent 
the best fitted power laws i.e. simple power law (SPL in red) and broken power law (BPL in green). 
The best-fitted model parameters calculated by the fitting procedure are summarized in Table 1.
A BPL model is favored to describe the $\gamma$-ray spectral shape over the 
SPL model with a difference of the logarithm of likelihood, $-2 \Delta L$ = 73.8 
which corresponds to a significance of the order of 10 $\sigma$ (see Table 1). So, we 
conclude that the GeV spectrum of the source is governed by a broken power law with break 
energy, $E_{break}$ = 3.5$\pm$0.05 GeV with power law indices, $\Gamma_1$ = 2.02$\pm$0.01 and 
$\Gamma_2$ = 2.40$\pm$0.04, respectively, below and above the break energy. The change in 
power law index ($\Delta \Gamma$) defined as $\Gamma_2 - \Gamma_1$ is 0.38$\pm$0.04.

\begin{figure*}
\includegraphics[scale=0.58,angle=-90, trim = 0 0 390 0, clip]{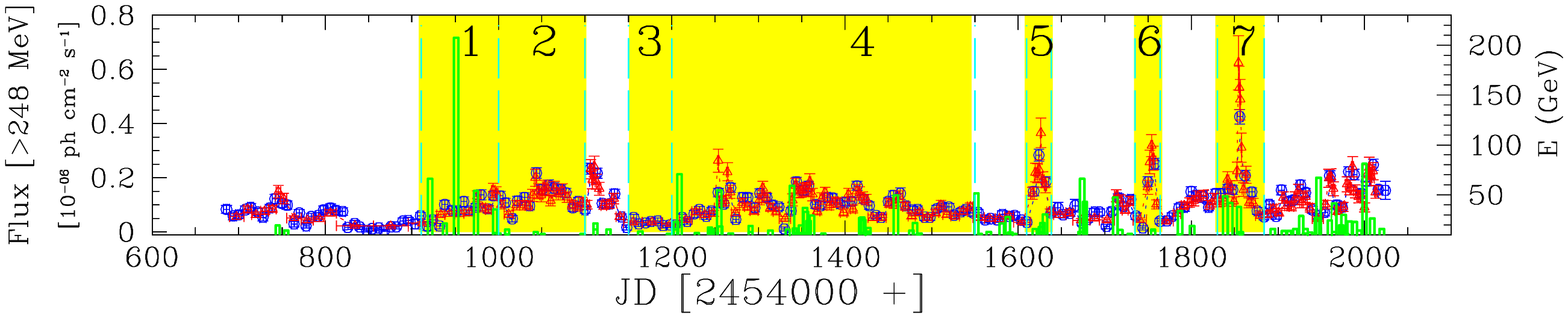}
\hspace{-0.2in}
\includegraphics[scale=0.2123,angle=0]{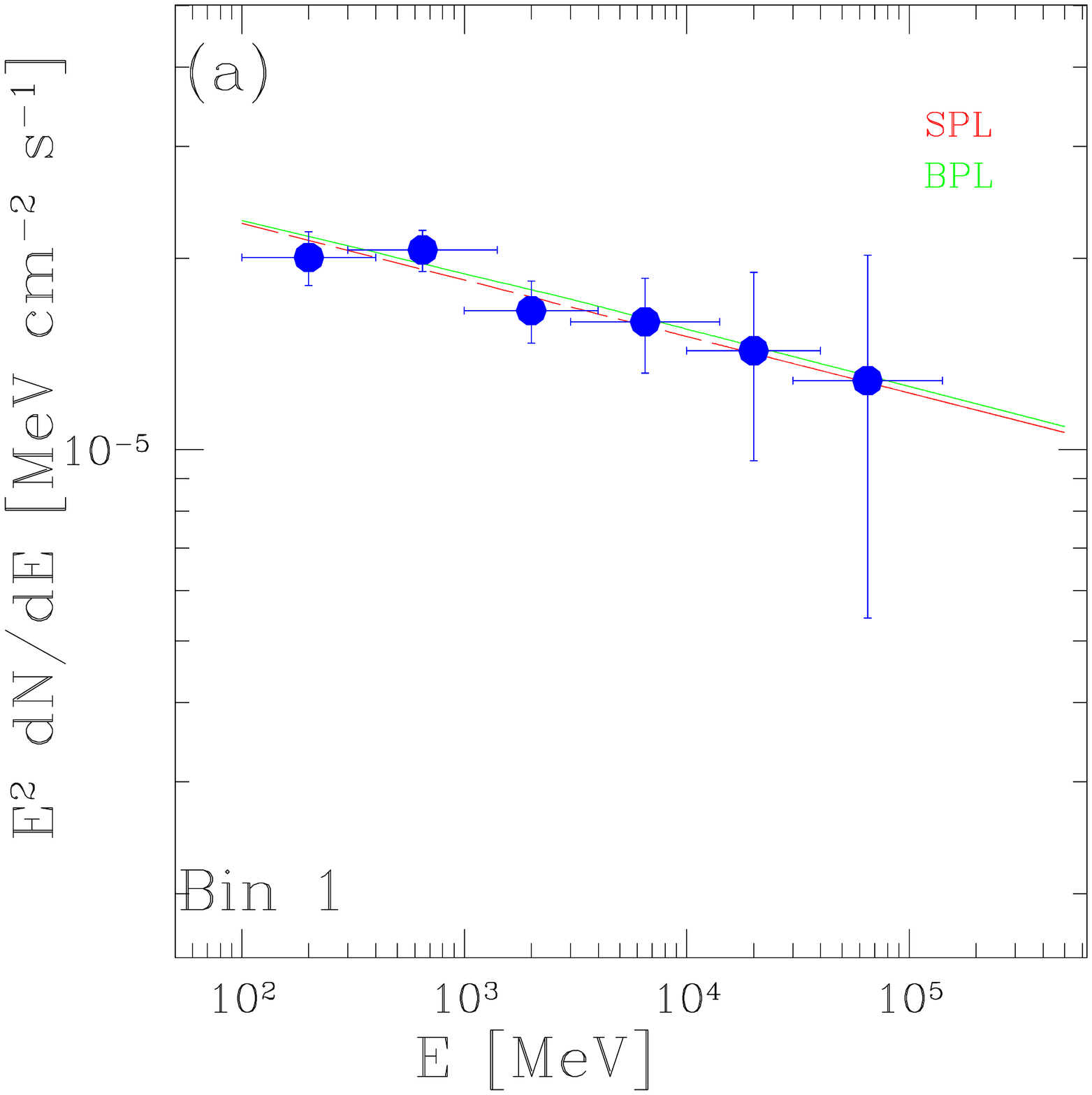}
\includegraphics[scale=0.2123,angle=0]{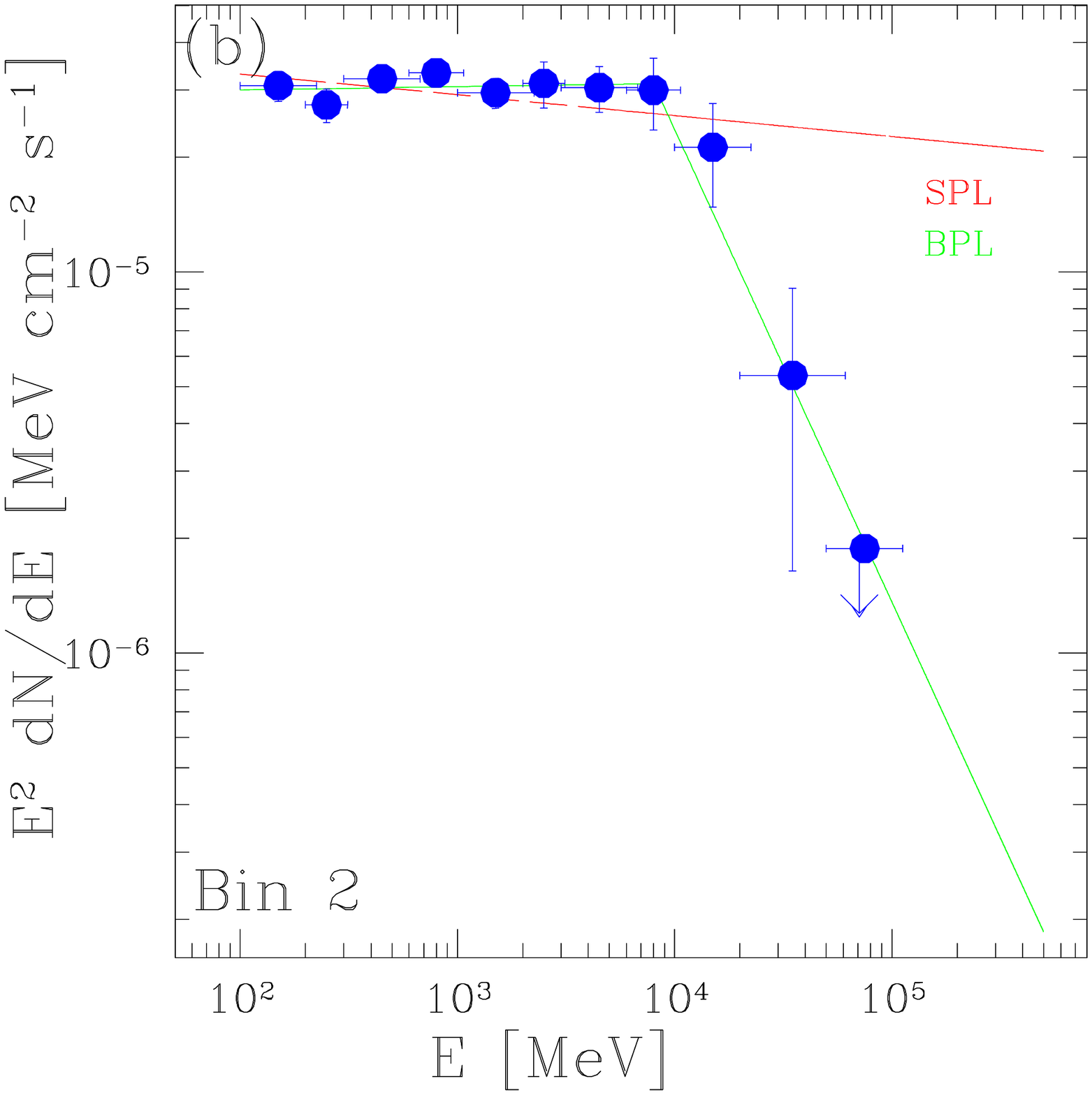}
\includegraphics[scale=0.2123,angle=0]{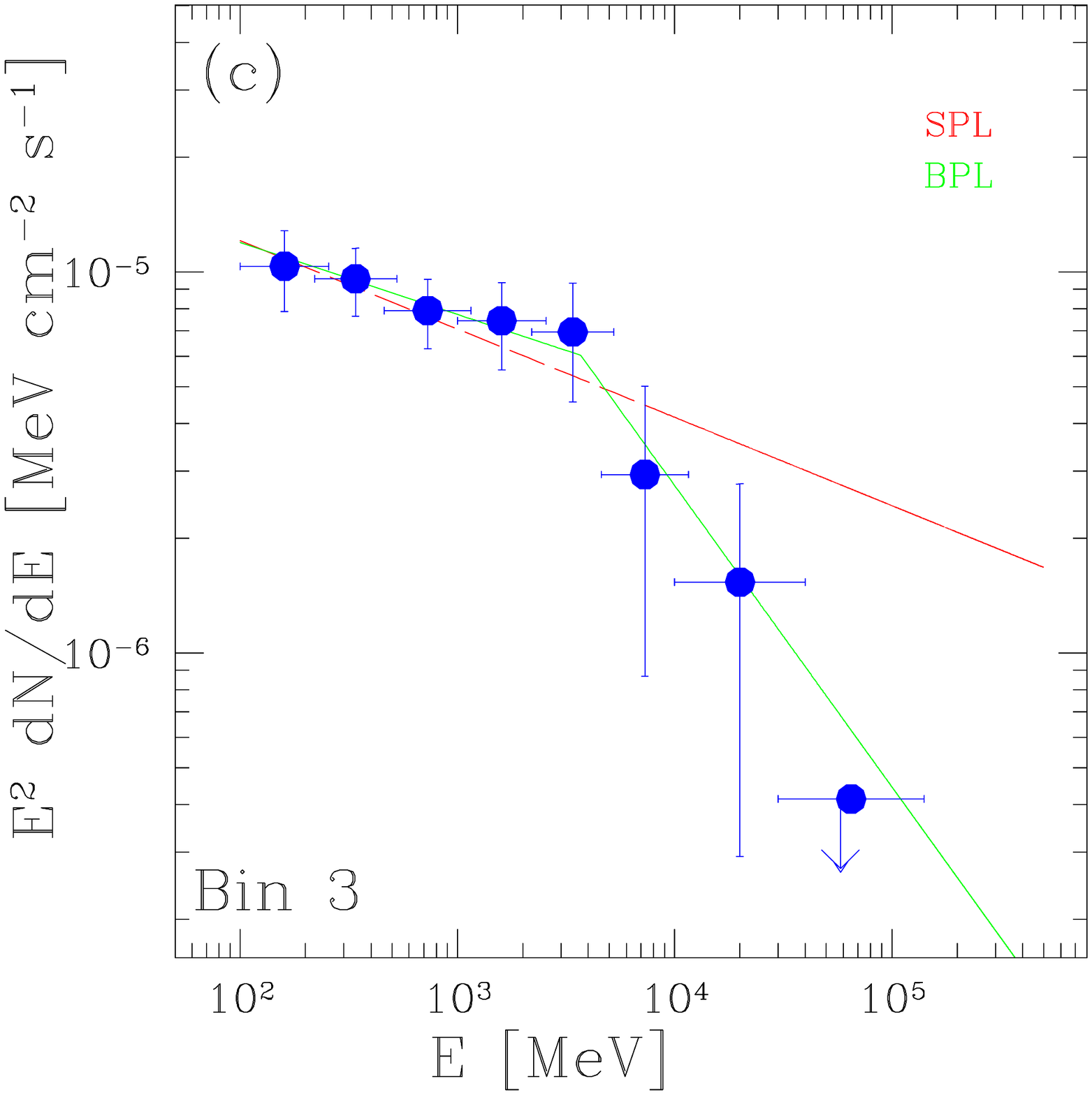}
\includegraphics[scale=0.2123,angle=0]{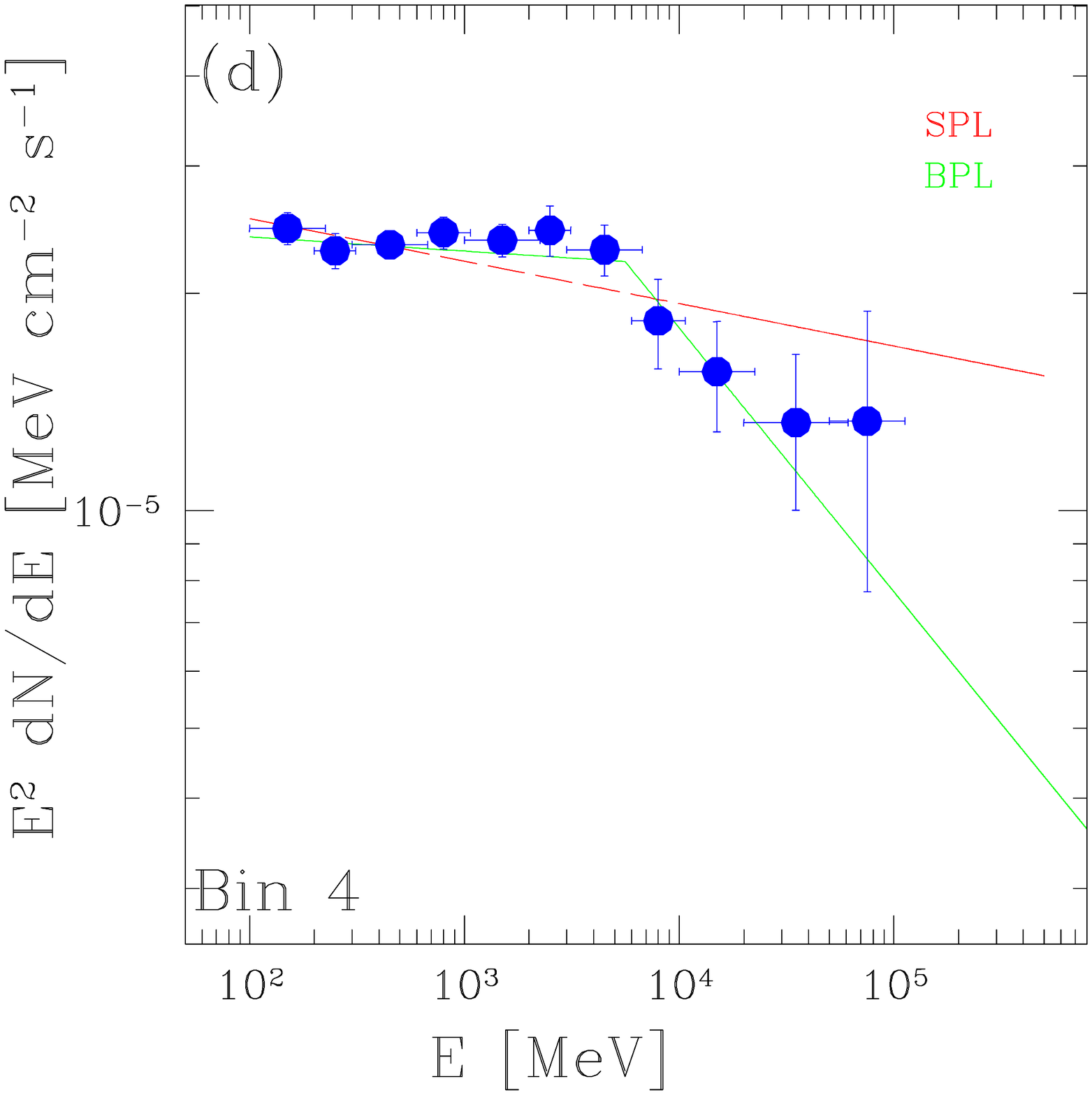}
\includegraphics[scale=0.2123,angle=0]{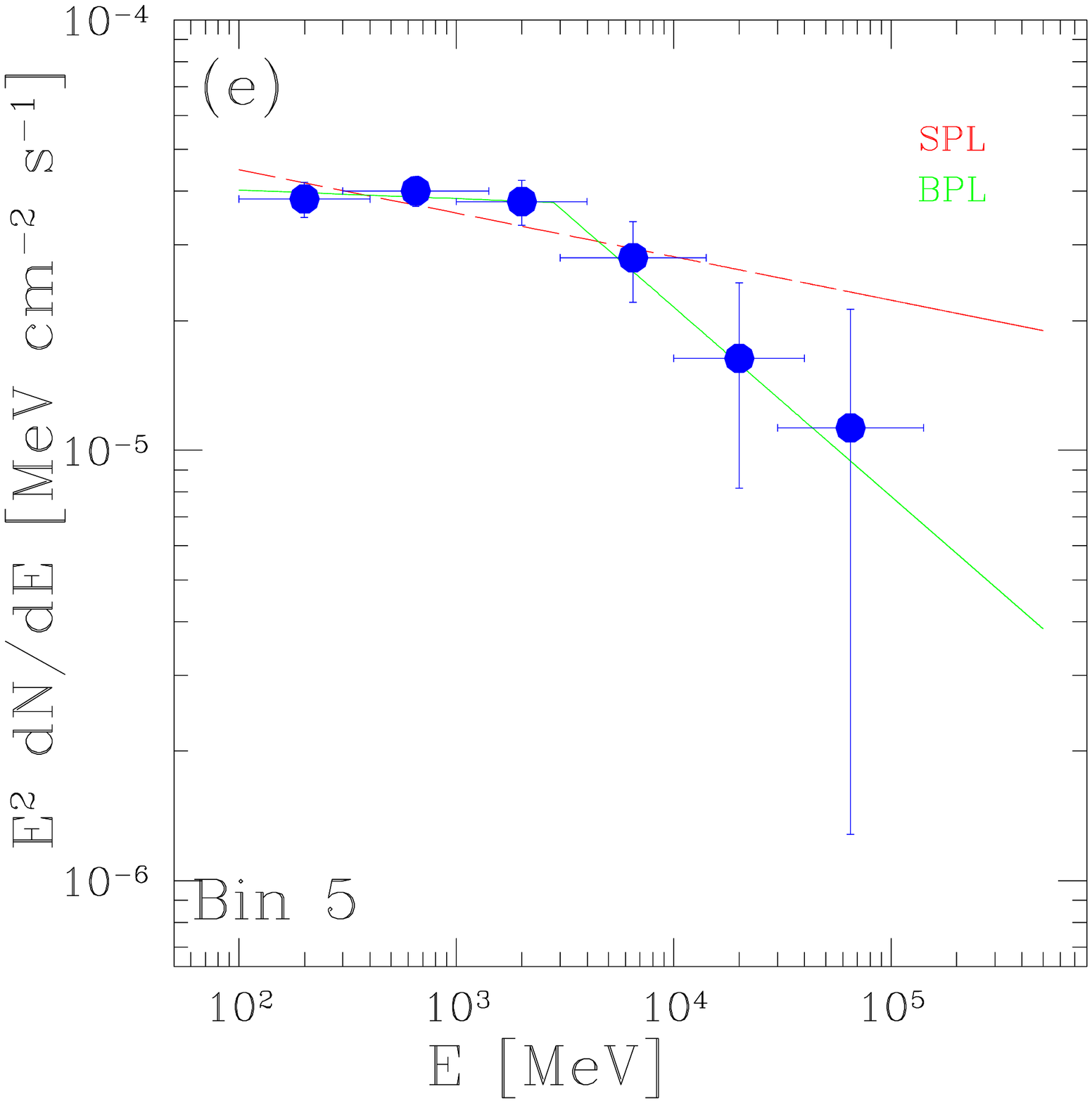}
\includegraphics[scale=0.2123,angle=0]{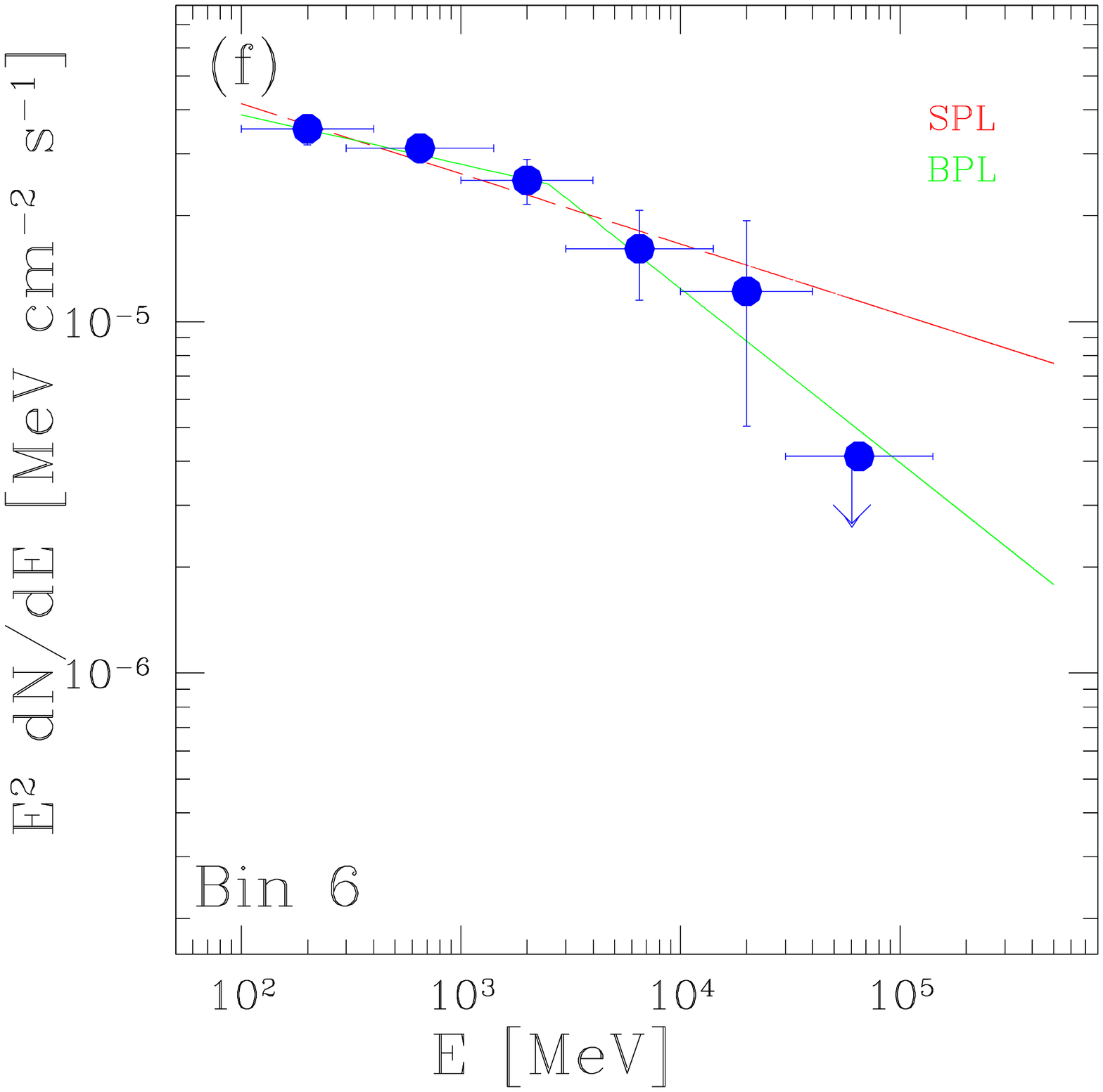}
\begin{minipage}{0.34\linewidth}
\includegraphics[scale=0.2123,angle=0]{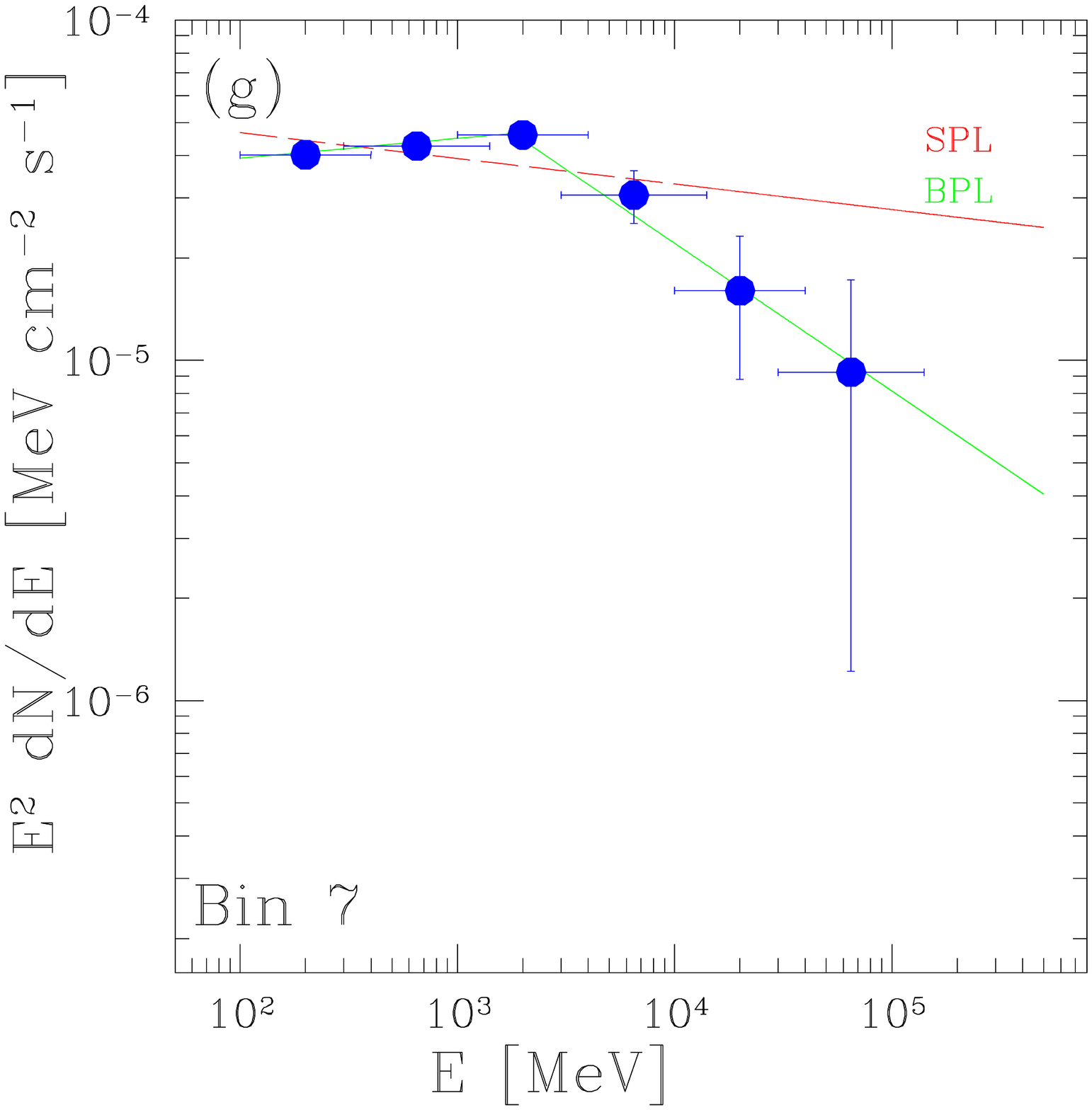}
\end{minipage}\hfill
\begin{minipage}[c]{0.63\linewidth}
   \caption{Top : Flux (E$>$248 MeV) light curve of the source over plotted with the arrival distribution of 
high energy photons E$>$20 GeV (same as in Fig. \ref{plot_weeklylc} (d)). The yellow area represents  
different activity periods of the source used to construct the GeV spectrum.
 (a)-(g) : Gamma-ray spectral energy distribution of S5 0716+714 during 
different activity states (shown in the top of the figure) along with the best fitted SPL (in green) and BPL (in red).  
              } 
\label{plot_sed_flares}
\end{minipage}
    \end{figure*}

It is very likely that the physical conditions within the emission region changes 
during different activity states. This motivates us to investigate the  
$\gamma$-ray spectrum for the individual flares. So, we compare the GeV spectra of the source 
during different activity states. Depending upon the 
flux variability state and the distribution of high energy photons (E$>$20 GeV), we construct   
the GeV spectrum of the source over seven different periods shown in the top of 
Fig. \ref{plot_sed_flares}. It is important to note that the spectral bins are not at equal time 
widths. We consider the following periods :  \\
{\bf Bin1} [JD' = 911-1000] : Flux is rising and a bunch of high energy 
(E$>$20 GeV) photons arrive during this period. \\
{\bf Bin2}  [JD' = 1000-1100] :  Flux level is high and no high energy photons arrive during this period. \\
{\bf Bin3}  [JD' = 1150-1200] : Flux level is very low and no high energy (E$>$15 GeV) photons arrive during this period. \\
{\bf Bin4} [JD' = 1200-1550]  : The source exhibits moderate level flux activity with a random distribution of 
arrival times of high energy photons. \\
{\bf Bin5}  [JD' = 1610-1638] :  A rapid flare with fewer high energy photons. \\
{\bf Bin6}  [JD' = 1735-1764] : A rapid flare with no high energy photons. \\
{\bf Bin7}  [JD' = 1840:1884] : The highest peaking flare with fewer high energy photons.

Fig. \ref{plot_sed_flares} (a)-(g) shows 
the individual GeV spectra of the source over these time bins. Here the blue symbols represent the spectral points 
constructed through a SPL fitting over the equally spaced logarithmic energy bins and the solid curves show 
the best fitted power law distributions.  The fitted parameters with the SPL and BPL models are given in Table \ref{tab_para}. 
The difference of the logarithm of likelihood $-2 \Delta L$ 
is given in the second last column of Table \ref{tab_para} with a significance level by which the BPL model is preferred 
over the SPL model  in the last column.  We find that a broken power law model is favored to describe the $\gamma$-ray 
spectral shape over the SPL model for all the time bins except Bin1. 
 For this bin, the estimated value of $2 \Delta L$ = 0.46, which corresponds to a significance level lower than 1 $\sigma$. 
Thus, the BPL model is not a better fit of the data than SPL. 
So, we do not find any clear break in GeV spectrum of the source for Bin1, rather,  a 
SPL model better describes the spectrum (see Fig. \ref{plot_sed_flares} (a)). 
Interestingly, the $\gamma$-ray flux of the source is rising and a bunch of high energy photons are observed 
over this period. In fact, the 207 GeV photon arrives during this period.  For the remaining time bins, the change 
in spectral index ($\Delta \Gamma$) below and above the break energy is listed in the column 8 of 
Table \ref{tab_para}. We found that $\Delta \Gamma$ varies between 0.38$\pm$0.02 to 1.14$\pm$0.40 over 
the different activity states of the source.

 \begin{figure*}
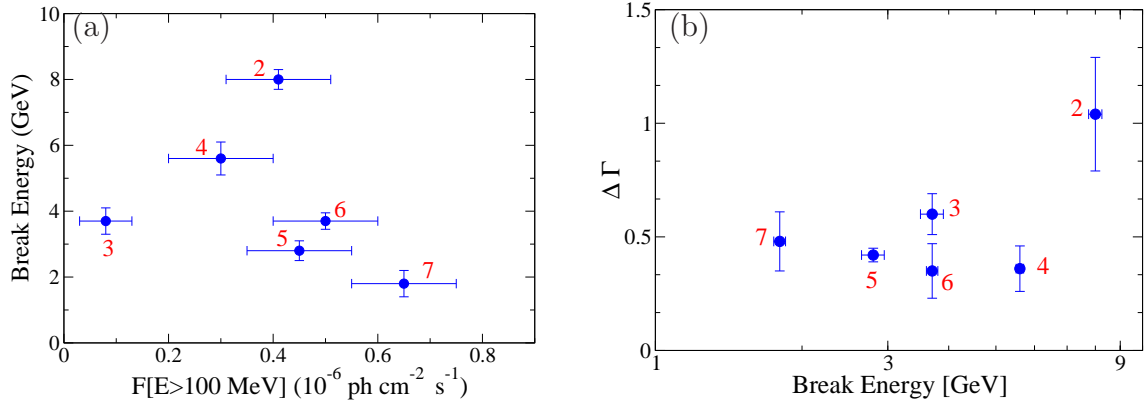

\includegraphics[scale=0.29,angle=0]{fig16.eps}
\put(-176,138){(a)}
\hspace{0.3in}
\includegraphics[scale=0.3,angle=0]{fig17.eps}
\put(-180,138){(b)}
   \caption{ (a) : Break Energy ($E_{Break}$) plotted as a function of flux for the different activity 
periods considered in Fig. 8 (b) to (g). (b) : Change of the spectral slope $\Delta \Gamma$ 
as a function of the break energy in the spectrum.               } 
\label{plot_Ebrk}
    \end{figure*}

 \begin{figure}
\includegraphics[scale=0.3,angle=0]{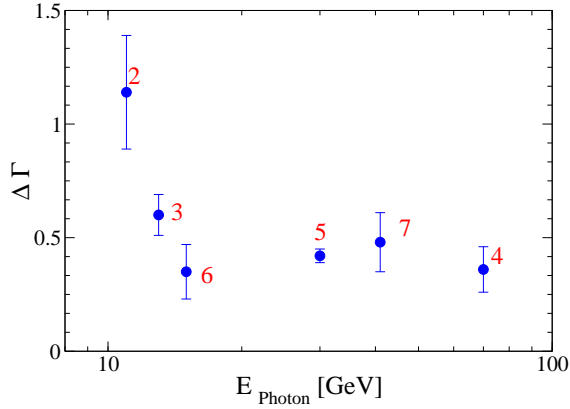}
   \caption{ Change in $\Delta \Gamma$ plotted as a function of E$_{Photon}$ for the different activity 
periods considered in Fig. \ref{plot_sed_flares} (b) to (g). E$_{Photon}$ is the energy corresponding to the 
highest energy detected photon.               } 
\label{plot_Eph}
    \end{figure}

\begin{table*}
\scriptsize
\caption{ Parameters of fitted power laws }
\begin{tabular}{c c c c c c c c c c } \hline 
Bin& JD'  & $F_{100}$   & Model&$\Gamma$/$\Gamma_1$  & $\Gamma_2$  & $E_{Break}$ &$\Delta \Gamma$ & $-2\Delta L$ & Significance  \\   
   &[JD-2454000] &($10^{-6} ph ~cm^{-2} s^{-1}$)&     &    &      & (GeV)          &         &        \\\hline
total& 680-2022 &              &SPL &2.09$\pm$0.01  &              &              &           &       &                   \\
     &          &              &BPL &2.02$\pm$0.01  &2.40$\pm$0.04 &3.50$\pm$0.05 &0.38$\pm$0.04  & 73.80 &$> 10 \sigma$\\\hline
1    & 911-1000 &0.21$\pm$0.10 &SPL &2.08$\pm$0.04  &              &              &               &       &             \\ 
     &          &              &BPL &2.11$\pm$0.05  &2.05$\pm$0.04 &3.00$\pm$0.25 &-0.06$\pm$0.06 & 0.46  &$< 1 \sigma$       \\\hline
2    & 1000-1100&0.41$\pm$0.11 &SPL &2.05$\pm$0.02  &              &              &               &       &             \\
     &          &              &BPL &1.99$\pm$0.03  &3.13$\pm$0.40 &8.00$\pm$0.25 &1.14$\pm$0.40  & 16.90 &$> 3 \sigma$ \\\hline
3    & 1150-1200&0.08$\pm$0.04 &SPL &2.23$\pm$0.09  &              &              &               &       &             \\
     &          &              &BPL &2.19$\pm$0.08  &2.79$\pm$0.50 &3.70$\pm$0.20 &0.60$\pm$0.09  & 13.22 &$> 3 \sigma$ \\\hline
4    & 1200-1550&0.27$\pm$0.10 &SPL &2.05$\pm$0.01  &              &              &               &       &             \\
     &          &              &BPL &2.01$\pm$0.03  &2.37$\pm$0.01 &5.6$\pm$0.10  &0.36$\pm$0.10  & 11.90 &$> 3 \sigma$ \\\hline
5    & 1610-1638&0.45$\pm$0.12 &SPL &2.10$\pm$0.04  &              &              &               &       &             \\
     &          &              &BPL &2.02$\pm$0.02  &2.44$\pm$0.03 &2.80$\pm$0.15 &0.42$\pm$0.03  & 11.72 &$ 3 \sigma$  \\\hline
6    & 1735-1764&0.51$\pm$0.11 &SPL &2.19$\pm$0.05  &              &              &               &       &             \\
     &          &              &BPL &2.14$\pm$0.07  &2.49$\pm$0.14 &3.70$\pm$0.10 &0.35$\pm$0.12  & 9.80  &$> 3 \sigma$    \\\hline
7    & 1840-1884&0.68$\pm$0.15 &SPL &2.07$\pm$0.03  &              &              &               &       &             \\
     &          &              &BPL &1.94$\pm$0.05  &2.42$\pm$0.13 &1.80$\pm$0.05 &0.48$\pm$0.13  & 10.52 &$> 3 \sigma$    \\\hline
\end{tabular} \\
$\Delta L$ is the difference of the -log(likelihood) value of BPL with respect to SPL.   
\label{tab_para}
\end{table*}

The variation of the break energy, $E_{break}$ with flux during the different activity states is displayed in 
Fig. \ref{plot_Ebrk} (a),  which does not show any systematic evolution of the break energy as a function of the flux 
variations.  The formal correlation statistics also does not reveal a significant correlation of the break energy w.r.t. the flux 
variations. Formally, we obtain the following correlation coefficient and significance : $r_P$ = -0.34 and 
55$\%$ confidence level ($r_P$ being the linear Pearson correlation coefficient). Similar to other Fermi blazars, 
we find no systematic variation of $E_{Break}$ as a function of the photon flux variations in S5 0716+714.

The variation of change in spectral slope ($\Delta \Gamma$) as a function of the break energy (E$_{Break}$) 
is shown in Fig. \ref{plot_Ebrk} (b). Again, we do not see any systematic variation in $\Delta \Gamma$ w.r.t. 
E$_{Break}$. Although, $\Delta \Gamma$ ($>$1) is higher for the higher break energy ($\sim$8 GeV), but, for the rest, 
$\Delta \Gamma$ remains almost constant with a decrease in  E$_{Break}$.

As we see in Fig. \ref{plot_sed_flares}, no spectral break is observed during the arrival period  of the highest energy 
photon (Bin 1). Likewise, $\Delta \Gamma$ is maximum for Bin 2 (no high energy 
photons arrived during this period).  This indicates a possible connection between the energy of highest detected GeV photon, 
$E_{Photon}$ and the spectral break parameters, as expected.  Fig. \ref{plot_Eph} 
shows the variation of $\Delta \Gamma$  as a function of $E_{Photon}$. As we see here, $\Delta \Gamma$ decreases with an increase 
in $E_{Photon}$. The correlation statistics reveal a significant 
correlation of $\Delta \Gamma$ w.r.t. $E_{Photon}$. Formally, we obtain the following correlation coefficient and 
confidence level, for $\Delta \Gamma$ vs $E_{Photon}$ : $r_P$ = -0.74 and 95$\%$ confidence level. 
Concluding this section, we can say that the detection of the high energy photons with energy 
$E_{Photon}$ is seems to be correlated with the spectral break parameter, $\Delta \Gamma$.

\section{Discussion} 
The GeV spectral breaks seen in many bright Fermi blazars lie within a few GeVs. The origin of these spectral breaks 
has generated considerable theoretical interest, and is still controversial. Among the most likely scenarios,  
the absorption of $\gamma$-rays via photon-photon pair production on He II Lyman recombination continuum and 
lines  within the broad line region \citep[e.g.][and references therein]{poutanen2010, tanaka2011} may be 
responsible for the observed breaks.  
Spectral breaks in the bright Fermi FSRQs like 3C 454.3, 3C 279, 
PKS 1510-089, 4C +21.35 etc. were interpreted using these scenarios. 
 $\gamma \gamma$ absorption by full BLR is also proposed as one reasonable possibility \citep{senturk2011}. 
The $\gamma$-ray emitting region must be located deep within the BLR for this model to work.

Alternatively, the GeV spectral breaks could also be explained by a combination of two Compton-scattered 
components, for example, by Compton scattering of the disk and BLR radiation as proposed by \citet{finke2010}. 
They explore this possibility to model the spectral breaks in FSRQ 3C 454.3. 
The combined external Compton and synchrotron self-Compton components may also explain these spectral breaks.
A further explanation invokes an intrinsic origin of the spectral breaks. The change in spectral index below and 
above the break of order  0.5 is expected from the typical ``cooling break" associated with radiative 
losses \citep{abdo2009}. The observed softening may instead be due to an intrinsic decline or break in the particle 
distribution as well.

We found that the change in spectral slope ($\Delta \Gamma$) above and below the break energy varies 
between 0.4 to 1.14. 
The estimated $\Delta \Gamma$ values for S5 0716+714 does not favor the standard radiative cooling models that 
predict a spectral break of 0.5 units.
It is also difficult to reconcile the constancy of the break energy w.r.t. the flux variations within the 
``cooling break" scenario.  Furthermore, this scenario failed to explain the absence of break for Bin 1. 
From this we conclude that the observed spectral breaks in 0716+714 are unlikely to have 
an intrinsic origin associated with the radiative cooling.

The spectral break in FSRQ 3C 454.3 is reproduced by a combination of two components, namely, the 
Compton-scattered disk and broad-line region (BLR) radiations \citep{finke2010}. But, as for BL Lacs,  
the jet radiation completely outshines the disk emission. So, the disk emission contribution seems to be 
negligible here, although, we can not exclude it completely. A further possibility is the combination of 
SSC and EC components. 
Modeling the broadband SEDs of the source over different time bins during the course of {\it Fermi/LAT} 
observations (see Rani et al. 2013 for details), we found that a model including an external Compton component 
generally does a better job in reproducing the entire SED with an external radiation field dominated 
by Ly-$\alpha$ from a putative broad line region (BLR). There we found that the radiation field energy 
density of this external field varies between 10$^{-(6 - 5)}$ ergs cm$^{-3}$, which is a factor of $\sim$1000 
lower than what we expect for a typical quasar.  
Such values are not unreasonable for BL Lac type objects and since S5 0716+714
is known to exhibit  a featureless optical spectrum. A low energy density of the BLR
appears in accordance with the non-detection of emission line. However, the non-existence
of a spectral break for `Bin 1' (see Fig. \ref{plot_sed_flares}) cannot be easily explained within this scenario.

The spectral breaks within few GeVs are well described by $\gamma$-ray absorption within the broad line region. 
The observed spectral breaks due to absorption within the broad line 
region constrain the location of the $\gamma$-ray emission region. It implies that $\gamma$-rays are produced 
within the BLR region i.e. within a few parsec distance from the central engine.  The $\gamma$-ray photons 
originating outside of the BLR region are unlikely to be absorbed as a result of $\gamma \gamma$ absorption and hence do 
not show any break in the GeV spectrum. Likewise the chances of detection of high energy photons will also be 
higher. So, it can be convincingly argued that the detection of many high energy photons and 
the absence of a spectral break for Bin1 is due to lower $\gamma \gamma$ absorption.

Alternatively, external Compton scattering of IR photons from a dusty torus offers an
alternative explanation for the observed spectral breaks. Given the fact that 0716+714
is also detected at TeV energies by MAGIC \citep{anderhub2009}, the frequency of the target
photons for the inverse Compton up-scattering in the Thompson regime 
should be less than 10$^{14}$ Hz \citep{sahayanathan2012}. If the scattering would be in the KN regime, a steeper
photon index should be seen, which is in disagreement of the observed hard TeV spectra.
However, we note that the lack of any excess IR detection expected from a torus, does not necessarily
rule out its existence, owing to the large amount of relativistic Doppler-boosting of the core region and
the resulting strong dominance of the non-thermal emission. In this context the physical nature
of the observed spectral break in the GeV/TeV spectrum still remains open and poses a 
challenge for future theoretical modeling.

The impact of the geometry of the broad line region on the expected absorption, through the $\gamma \gamma$ process was 
recently discussed by \citet{tavecchio2012}.  They argued that a correlated variation in $\Delta \Gamma$ and  $E_{Break}$ 
is expected for an ``open" geometry of the BLR. 
However, if the BLR is ``closed" the break energy does not change as long as the emission occurs within the BLR, but $\Delta \Gamma$ 
decreases as the emission region moves away from the central engine.  
For 0716+714, we do no find any correlated variation between $\Delta \Gamma$ and  $E_{Break}$, although, 
both changes from state to state. This rules out the possibility of an ``open" BLR geometry and most probably the emission 
region in the source is not  located at a fixed distance from the black hole. 
We also notice that the GeV spectrum constructed during the period of detection of the highest energy (207 GeV) photon does not 
show any spectral break, while, the spectral break parameters ($\Delta \Gamma$ and  $E_{Break}$) are maximum for the spectrum 
constructed over the period where the detection of high energy (E$>$20 GeV) photons is low. 
A significant correlation between $\Delta \Gamma$ and $E_{photon}$ with  a decreasing $\Delta \Gamma$ for an increasing  
$E_{photon}$ is a signature of varying opacity.

\section{Conclusions} 
The continuous monitoring in the high-energy $\gamma$-ray band by {\it Fermi/LAT} allows us to investigate the 
GeV flux and spectral variability of the BL Lac object S5 0716+714. The source displays prominent flaring 
activity during this period reaching as high as $\sim$ 1.5 $\times$ 10$^{-6}$ ph cm$^{-2}$ s$^{-1}$ above 100 MeV. 
The source exhibits two different modes of variability : (1) the slow and modest level flux variability, and (2) the 
rapid flares. The estimated variability 
timescale for the two modes of variability are 140$\pm$5 and 75$\pm$5 days, respectively. 
The highest recorded photon for the source arrives at ``207 GeV", and is observed during the
rising part of the first flare at JD = 2454951. Similar to other Fermi blazars, no significant correlation 
between the flux and photon index has been measured in the source, rather the flux variations are characterized 
by a weak spectral hardening.  A more detailed discussion 
of the broadband flaring activity with emission models  will be given in Rani et al. (2012).

The 3.8 year averaged $\gamma$-ray spectral shape above 100 MeV clearly deviates from a 
single power law. A broken-power law model yields a break energy within a few GeV range. During 
different activity states of the source, the spectral break energy does not follow any systematic trend 
w.r.t the photon flux variations. Such a behavior is 
similar to that observed in other bright Fermi blazars. The combination of non-simultaneous GeV-TeV 
spectrum of the source shows absorption 
like features between 10-100 GeV \citep{senturk2011}. More simultaneous GeV-TeV observations are required 
to check this. A continuous TeV monitoring of the source during will 
shed more light on it. 
This study has highlighted some possible explanations for the origin of GeV spectral breaks in BL Lac 
S5 0716+714. Following our analysis, we argued in favor of $\gamma \gamma$ absorption. 
Still, we can not rule out a combination of two or more Compton scattered components which could also lead such breaks.   
We address all these key questions for future study. \\

\noindent
{\bf Acknowledgments} 

\noindent
The {\it Fermi-LAT} Collaboration acknowledges the generous support of a number of agencies 
and institutes that have supported the {\it Fermi-LAT} Collaboration. These include the National 
Aeronautics and Space Administration and the Department of Energy in the United States, the 
Commissariat \`a l'Energie Atomique and the Centre National de la Recherche Scientifique / Institut 
National de Physique Nucl\'eaire et de Physique des Particules in France, the Agenzia Spaziale 
Italiana and the Istituto Nazionale di Fisica Nucleare in Italy, the Ministry of Education, 
Culture, Sports, Science and Technology (MEXT), High Energy Accelerator Research Organization 
(KEK) and Japan Aerospace Exploration Agency (JAXA) in Japan, and the K.\ A.\ Wallenberg 
Foundation, the Swedish Research Council and the Swedish National Space Board in Sweden. 
BR gratefully acknowledges the travel support from the COSPAR Capacity-Building 
Workshop fellowship program. BR was supported for this research through a stipend from the International Max 
Planck Research School (IMPRS) for Astronomy and Astrophysics at the Universities of Bonn and Cologne.








\begin{thebibliography}{24}
\expandafter\ifx\csname natexlab\endcsname\relax\def\natexlab#1{#1}\fi

\bibitem[{{Abdo} {et~al.}(2009){Abdo}, {Ackermann}, {Ajello}, {Atwood},
  {Axelsson}, {Baldini}, {Ballet}, {Band}, {Barbiellini}, {Bastieri}, \&
  et~al.}]{abdo2009}
{Abdo}, A.~A., {Ackermann}, M., {Ajello}, M., {et~al.} 2009, ApJS, 183, 46

\bibitem[{{Abdo} {et~al.}(2010){Abdo}, {Ackermann}, {Ajello}, {Atwood},
  {Axelsson}, {Baldini}, {Ballet}, {Barbiellini}, {Bastieri}, {Bechtol},
  {Bellazzini}, {Berenji}, {Blandford}, {Bloom}, {Bonamente}, {Borgland},
  {Bouvier}, {Bregeon}, {Brez}, {Brigida}, {Bruel}, {Burnett}, {Buson},
  {Caliandro}, {Cameron}, {Caraveo}, {Carrigan}, {Casandjian}, {Cavazzuti},
  {Cecchi}, {{\c C}elik}, {Charles}, {Chekhtman}, {Cheung}, {Chiang},
  {Ciprini}, {Claus}, {Cohen-Tanugi}, {Conrad}, {Cutini}, {Dermer}, {de
  Angelis}, {de Palma}, {Digel}, {Silva}, {Drell}, {Dubois}, {Dumora},
  {Farnier}, {Favuzzi}, {Fegan}, {Focke}, {Fortin}, {Frailis}, {Fukazawa},
  {Funk}, {Fusco}, {Gargano}, {Gasparrini}, {Gehrels}, {Germani}, {Giebels},
  {Giglietto}, {Giommi}, {Giordano}, {Glanzman}, {Godfrey}, {Grenier},
  {Grondin}, {Grove}, {Guillemot}, {Guiriec}, {Harding}, {Hartman},
  {Hayashida}, {Hays}, {Healey}, {Horan}, {Hughes}, {Jackson},
  {J{\'o}hannesson}, {Johnson}, {Johnson}, {Kamae}, {Katagiri}, {Kataoka},
  {Kawai}, {Kerr}, {Kn{\"o}dlseder}, {Kuss}, {Lande}, {Latronico},
  {Lemoine-Goumard}, {Longo}, {Loparco}, {Lott}, {Lovellette}, {Lubrano},
  {Madejski}, {Makeev}, {Mazziotta}, {McConville}, {McEnery}, {Meurer},
  {Michelson}, {Mitthumsiri}, {Mizuno}, {Moiseev}, {Monte}, {Monzani},
  {Morselli}, {Moskalenko}, {Murgia}, {Nolan}, {Norris}, {Nuss}, {Ohsugi},
  {Omodei}, {Orlando}, {Ormes}, {Paneque}, {Panetta}, {Parent}, {Pelassa},
  {Pepe}, {Persic}, {Pesce-Rollins}, {Piron}, {Porter}, {Rain{\`o}}, {Rando},
  {Razzano}, {Reimer}, {Reimer}, {Reposeur}, {Ritz}, {Rochester}, {Rodriguez},
  {Romani}, {Roth}, {Ryde}, {Sadrozinski}, {Sanchez}, {Sander}, {Saz
  Parkinson}, {Scargle}, {Sgr{\`o}}, {Siskind}, {Smith}, {Smith}, {Spandre},
  {Spinelli}, {Strickman}, {Suson}, {Tajima}, {Takahashi}, {Takahashi},
  {Tanaka}, {Thayer}, {Thayer}, {Thompson}, {Tibaldo}, {Torres}, {Tosti},
  {Tramacere}, {Uchiyama}, {Usher}, {Vasileiou}, {Vilchez}, {Villata},
  {Vitale}, {Waite}, {Wang}, {Winer}, {Wood}, {Ylinen}, \&
  {Ziegler}}]{abdo2010LBAS}
{Abdo}, A.~A., {Ackermann}, M., {Ajello}, M., {et~al.} 2010, ApJ, 710, 1271

\bibitem[{{Ackermann} {et~al.}(2011){Ackermann}, {Ajello}, {Allafort},
  {Antolini}, {Atwood}, {Axelsson}, {Baldini}, {Ballet}, {Barbiellini},
  {Bastieri}, {Bechtol}, {Bellazzini}, {Berenji}, {Blandford}, {Bloom},
  {Bonamente}, {Borgland}, {Bottacini}, {Bouvier}, {Bregeon}, {Brigida},
  {Bruel}, {Buehler}, {Burnett}, {Buson}, {Caliandro}, {Cameron}, {Caraveo},
  {Casandjian}, {Cavazzuti}, {Cecchi}, {Charles}, {Cheung}, {Chiang},
  {Ciprini}, {Claus}, {Cohen-Tanugi}, {Conrad}, {Costamante}, {Cutini}, {de
  Angelis}, {de Palma}, {Dermer}, {Digel}, {Silva}, {Drell}, {Dubois},
  {Escande}, {Favuzzi}, {Fegan}, {Ferrara}, {Finke}, {Focke}, {Fortin},
  {Frailis}, {Fukazawa}, {Funk}, {Fusco}, {Gargano}, {Gasparrini}, {Gehrels},
  {Germani}, {Giebels}, {Giglietto}, {Giommi}, {Giordano}, {Giroletti},
  {Glanzman}, {Godfrey}, {Grenier}, {Grove}, {Guiriec}, {Gustafsson},
  {Hadasch}, {Hayashida}, {Hays}, {Healey}, {Horan}, {Hou}, {Hughes},
  {Iafrate}, {J{\'o}hannesson}, {Johnson}, {Johnson}, {Kamae}, {Katagiri},
  {Kataoka}, {Kn{\"o}dlseder}, {Kuss}, {Lande}, {Larsson}, {Latronico},
  {Longo}, {Loparco}, {Lott}, {Lovellette}, {Lubrano}, {Madejski}, {Mazziotta},
  {McConville}, {McEnery}, {Michelson}, {Mitthumsiri}, {Mizuno}, {Moiseev},
  {Monte}, {Monzani}, {Moretti}, {Morselli}, {Moskalenko}, {Murgia},
  {Nakamori}, {Naumann-Godo}, {Nolan}, {Norris}, {Nuss}, {Ohno}, {Ohsugi},
  {Okumura}, {Omodei}, {Orienti}, {Orlando}, {Ormes}, {Ozaki}, {Paneque},
  {Parent}, {Pesce-Rollins}, {Pierbattista}, {Piranomonte}, {Piron}, {Pivato},
  {Porter}, {Rain{\`o}}, {Rando}, {Razzano}, {Razzaque}, {Reimer}, {Reimer},
  {Ritz}, {Rochester}, {Romani}, {Roth}, {Sanchez}, {Sbarra}, {Scargle},
  {Schalk}, {Sgr{\`o}}, {Shaw}, {Siskind}, {Spandre}, {Spinelli}, {Strong},
  {Suson}, {Tajima}, {Takahashi}, {Takahashi}, {Tanaka}, {Thayer}, {Thayer},
  {Thompson}, {Tibaldo}, {Tinivella}, {Torres}, {Tosti}, {Troja}, {Uchiyama},
  {Vandenbroucke}, {Vasileiou}, {Vianello}, {Vitale}, {Waite}, {Wallace},
  {Wang}, {Winer}, {Wood}, {Wood}, \& {Zimmer}}]{ackermann2011}
{Ackermann}, M., {Ajello}, M., {Allafort}, A., {et~al.} 2011, ApJ, 743, 171

\bibitem[{{Anderhub} {et~al.}(2009){Anderhub}, {Antonelli}, {Antoranz},
  {Backes}, {Baixeras}, {Balestra}, {Barrio}, {Bastieri}, {Becerra
  Gonz{\'a}lez}, {Becker}, {Bednarek}, {Berdyugin}, {Berger}, {Bernardini},
  {Biland}, {Bock}, {Bonnoli}, {Bordas}, {Borla Tridon}, {Bosch-Ramon}, {Bose},
  {Braun}, {Bretz}, {Britzger}, {Camara}, {Carmona}, {Carosi}, {Colin},
  {Commichau}, {Contreras}, {Cortina}, {Costado}, {Covino}, {Dazzi}, {De
  Angelis}, {de Cea del Pozo}, {De los Reyes}, {De Lotto}, {De Maria}, {De
  Sabata}, {Delgado Mendez}, {Dom{\'{\i}}nguez}, {Dominis Prester}, {Dorner},
  {Doro}, {Elsaesser}, {Errando}, {Ferenc}, {Fern{\'a}ndez}, {Firpo},
  {Fonseca}, {Font}, {Galante}, {Garc{\'{\i}}a L{\'o}pez}, {Garczarczyk},
  {Gaug}, {Godinovic}, {Goebel}, {Hadasch}, {Herrero}, {Hildebrand},
  {H{\"o}hne-M{\"o}nch}, {Hose}, {Hrupec}, {Hsu}, {Jogler}, {Klepser},
  {Kranich}, {La Barbera}, {Laille}, {Leonardo}, {Lindfors}, {Lombardi},
  {Longo}, {L{\'o}pez}, {Lorenz}, {Majumdar}, {Maneva}, {Mankuzhiyil},
  {Mannheim}, {Maraschi}, {Mariotti}, {Mart{\'{\i}}nez}, {Mazin}, {Meucci},
  {Miranda}, {Mirzoyan}, {Miyamoto}, {Mold{\'o}n}, {Moles}, {Moralejo},
  {Nieto}, {Nilsson}, {Ninkovic}, {Orito}, {Oya}, {Paoletti}, {Paredes},
  {Pasanen}, {Pascoli}, {Pauss}, {Pegna}, {Perez-Torres}, {Persic}, {Peruzzo},
  {Prada}, {Prandini}, {Puchades}, {Puljak}, {Reichardt}, {Rhode}, {Rib{\'o}},
  {Rico}, {Rissi}, {Robert}, {R{\"u}gamer}, {Saggion}, {Sainio}, {Saito},
  {Salvati}, {S{\'a}nchez-Conde}, {Satalecka}, {Scalzotto}, {Scapin},
  {Schweizer}, {Shayduk}, {Shore}, {Sierpowska-Bartosik}, {Sillanp{\"a}{\"a}},
  {Sitarek}, {Sobczynska}, {Spanier}, {Spiro}, {Stamerra}, {Stark}, {Suric},
  {Takalo}, {Tavecchio}, {Temnikov}, {Tescaro}, {Teshima}, {Torres}, {Turini},
  {Vankov}, {Wagner}, {Villforth}, {Zabalza}, {Zandanel}, {Zanin}, \&
  {Zapatero}}]{anderhub2009}
{Anderhub}, H., {Antonelli}, L.~A., {Antoranz}, P., {et~al.} 2009, ApJ, 704,
  L129

\bibitem[{{Chen} {et~al.}(2008){Chen}, {D'Ammando}, {Villata}, {Raiteri},
  {Tavani}, {Vittorini}, {Bulgarelli}, {Donnarumma}, {Ferrari}, {Giuliani},
  {Longo}, {Pacciani}, {Pucella}, {Vercellone}, {Argan}, {Barbiellini},
  {Boffelli}, {Caraveo}, {Carosati}, {Cattaneo}, {Cocco}, {Costa}, {Del Monte},
  {de Paris}, {Di Cocco}, {Evangelista}, {Feroci}, {Fiorini}, {Froysland},
  {Frutti}, {Fuschino}, {Galli}, {Gianotti}, {Kurtanidze}, {Labanti},
  {Lapshov}, {Larionov}, {Lazzarotto}, {Lipari}, {Marisaldi}, {Mastropietro},
  {Mereghetti}, {Morelli}, {Morselli}, {Pasanen}, {Pellizzoni}, {Perotti},
  {Picozza}, {Porrovecchio}, {Prest}, {Rapisarda}, {Rappoldi}, {Rubini},
  {Soffitta}, {Trifoglio}, {Trois}, {Vallazza}, {Zambra}, {Zanello}, {Cutini},
  {Gasparrini}, {Pittori}, {Santolamazza}, {Verrecchia}, {Giommi}, {Antonelli},
  {Colafrancesco}, \& {Salotti}}]{chen2008}
{Chen}, A.~W., {D'Ammando}, F., {Villata}, M., {et~al.} 2008, A\&A, 489, L37

\bibitem[{{Danforth} {et~al.}(2012){Danforth}, {Nalewajko}, {France}, \&
  {Keeney}}]{danforth2012}
{Danforth}, C.~W., {Nalewajko}, K., {France}, K., \& {Keeney}, B.~A. 2012,
  ArXiv e-prints

\bibitem[{{Ferrero} {et~al.}(2006){Ferrero}, {Wagner}, {Emmanoulopoulos}, \&
  {Ostorero}}]{ferrero2006}
{Ferrero}, E., {Wagner}, S.~J., {Emmanoulopoulos}, D., \& {Ostorero}, L. 2006,
  A\&A, 457, 133

\bibitem[{{Finke} \& {Dermer}(2010)}]{finke2010}
{Finke}, J.~D. \& {Dermer}, C.~D. 2010, ApJ, 714, L303

\bibitem[{{Foschini} {et~al.}(2006){Foschini}, {Tagliaferri}, {Pian},
  {Ghisellini}, {Treves}, {Maraschi}, {Tavecchio}, {Di Cocco}, \&
  {Rosen}}]{foschini2006}
{Foschini}, L., {Tagliaferri}, G., {Pian}, E., {et~al.} 2006, A\&A, 455, 871

\bibitem[{{Giommi} {et~al.}(1999){Giommi}, {Massaro}, {Chiappetti}, {Ferrara},
  {Ghisellini}, {Jang}, {Maesano}, {Miller}, {Montagni}, {Nesci}, {Padovani},
  {Perlman}, {Raiteri}, {Sclavi}, {Tagliaferri}, {Tosti}, \&
  {Villata}}]{giommi1999}
{Giommi}, P., {Massaro}, E., {Chiappetti}, L., {et~al.} 1999, A\&A, 351, 59

\bibitem[{{Gupta} {et~al.}(2012){Gupta}, {Krichbaum}, {Wiita}, {Rani},
  {Sokolovsky}, {Mohan}, {Mangalam}, {Marchili}, {Fuhrmann}, {Agudo}, {Bach},
  {Bachev}, {B{\"o}ttcher}, {Gabanyi}, {Gaur}, {Hawkins}, {Kimeridze},
  {Kurtanidze}, {Kurtanidze}, {Lee}, {Liu}, {McBreen}, {Nesci}, {Nestoras},
  {Nikolashvili}, {Ohlert}, {Palma}, {Peneva}, {Pursimo}, {Semkov},
  {Strigachev}, {Webb}, {Wiesemeyer}, \& {Zensus}}]{gupta2012}
{Gupta}, A.~C., {Krichbaum}, T.~P., {Wiita}, P.~J., {et~al.} 2012, MNRAS, 425,
  1357

\bibitem[{{Hartman} {et~al.}(1999){Hartman}, {Bertsch}, {Bloom}, {Chen},
  {Deines-Jones}, {Esposito}, {Fichtel}, {Friedlander}, {Hunter}, {McDonald},
  {Sreekumar}, {Thompson}, {Jones}, {Lin}, {Michelson}, {Nolan}, {Tompkins},
  {Kanbach}, {Mayer-Hasselwander}, {M{\"u}cke}, {Pohl}, {Reimer}, {Kniffen},
  {Schneid}, {von Montigny}, {Mukherjee}, \& {Dingus}}]{hartman1999}
{Hartman}, R.~C., {Bertsch}, D.~L., {Bloom}, S.~D., {et~al.} 1999, ApJS, 123,
  79

\bibitem[{{Lin} {et~al.}(1995){Lin}, {Bertsch}, {Dingus}, {Esposito},
  {Fichtel}, {Hartman}, {Hunter}, {Kanbach}, {Kniffen}, {Mayer-Hasselwander},
  {Michelson}, {von Montigny}, {Mukherjee}, {Nolan}, {Schneid}, {Sreekumar}, \&
  {Thompson}}]{lin1995}
{Lin}, Y.~C., {Bertsch}, D.~L., {Dingus}, B.~L., {et~al.} 1995, ApJ, 442, 96

\bibitem[{{Lott} {et~al.}(2012){Lott}, {Escande}, {Larsson}, \&
  {Ballet}}]{lott2012}
{Lott}, B., {Escande}, L., {Larsson}, S., \& {Ballet}, J. 2012, A\&A, 544, A6

\bibitem[{{Nandikotkur} {et~al.}(2007){Nandikotkur}, {Jahoda}, {Hartman},
  {Mukherjee}, {Sreekumar}, {B{\"o}ttcher}, {Sambruna}, \&
  {Swank}}]{nandikotkur2007}
{Nandikotkur}, G., {Jahoda}, K.~M., {Hartman}, R.~C., {et~al.} 2007, ApJ, 657,
  706

\bibitem[{{Nilsson} {et~al.}(2008){Nilsson}, {Pursimo}, {Sillanp{\"a}{\"a}},
  {Takalo}, \& {Lindfors}}]{nilsson2008}
{Nilsson}, K., {Pursimo}, T., {Sillanp{\"a}{\"a}}, A., {Takalo}, L.~O., \&
  {Lindfors}, E. 2008, A\&A, 487, L29

\bibitem[{{Poutanen} \& {Stern}(2010)}]{poutanen2010}
{Poutanen}, J. \& {Stern}, B. 2010, ApJ, 717, L118

\bibitem[{{Raiteri} {et~al.}(2003){Raiteri}, {Villata}, {Tosti}, {Nesci},
  {Massaro}, {Aller}, {Aller}, {Ter{\"a}sranta}, {Kurtanidze}, {Nikolashvili},
  {Ibrahimov}, {Papadakis}, {Krichbaum}, {Kraus}, {Witzel}, {Ungerechts},
  {Lisenfeld}, {Bach}, {Cim{\`o}}, {Ciprini}, {Fuhrmann}, {Kimeridze},
  {Lanteri}, {Maesano}, {Montagni}, {Nucciarelli}, \& {Ostorero}}]{raiteri2003}
{Raiteri}, C.~M., {Villata}, M., {Tosti}, G., {et~al.} 2003, A\&A, 402, 151

\bibitem[{{Rani} {et~al.}(2010){Rani}, {Gupta}, {Joshi}, {Ganesh}, \&
  {Wiita}}]{rani2010}
{Rani}, B., {Gupta}, A.~C., {Joshi}, U.~C., {Ganesh}, S., \& {Wiita}, P.~J.
  2010, ApJ, 719, L153

\bibitem[{{Rani} {et~al.}(2009){Rani}, {Wiita}, \& {Gupta}}]{rani2009}
{Rani}, B., {Wiita}, P.~J., \& {Gupta}, A.~C. 2009, ApJ, 696, 2170

\bibitem[Rani et al.(2013)]{rani2013} Rani, B., Krichbaum, 
T.~P., Fuhrmann, L., et al.\ 2013, arXiv:1301.7087 

\bibitem[Sahayanathan 
\& Godambe(2012)]{sahayanathan2012} Sahayanathan, S., \& Godambe, S.\ 2012, MNRAS, 419, 1660 

\bibitem[{{Senturk} {et~al.}(2011){Senturk}, {Errando}, {Boettcher}, {Coppi},
  {Mukherjee}, \& {Roustazadeh}}]{senturk2011}
{Senturk}, G.~D., {Errando}, M., {Boettcher}, M., {et~al.} 2011, ArXiv e-prints

\bibitem[{{Simonetti} {et~al.}(1985){Simonetti}, {Cordes}, \&
  {Heeschen}}]{simonetti1985}
{Simonetti}, J.~H., {Cordes}, J.~M., \& {Heeschen}, D.~S. 1985, ApJ, 296, 46

\bibitem[{{Tanaka} {et~al.}(2011){Tanaka}, {Stawarz}, {Thompson}, {D'Ammando},
  {Fegan}, {Lott}, {Wood}, {Cheung}, {Finke}, {Buson}, {Escande}, {Saito},
  {Ohno}, {Takahashi}, {Donato}, {Chiang}, {Giroletti}, {Schinzel}, {Iafrate},
  {Longo}, \& {Ciprini}}]{tanaka2011}
{Tanaka}, Y.~T., {Stawarz}, {\L}., {Thompson}, D.~J., {et~al.} 2011, ApJ, 733,
  19

\bibitem[{{Tavecchio} \& {Ghisellini}(2012)}]{tavecchio2012}
{Tavecchio}, F. \& {Ghisellini}, G. 2012, ArXiv e-prints

\end{thebibliography}
\end{document}